\documentclass[aps,prl,twocolumn,showpacs,superscriptaddress]{revtex4-1}
\usepackage{amsmath,amssymb}
\usepackage{graphics,graphicx,color}
\usepackage{dcolumn,bm}
\usepackage[normalem]{ulem}
\usepackage{tikz}
\usepackage{soul}
\usepackage[colorlinks=true,citecolor=blue,urlcolor=blue]{hyperref}
\usepackage{blkarray}

\def\[{\left[}
\def\]{\right]}
\def\({\left(}
\def\){\right)}
\def\be{\begin{equation}}
\def\ee{\end{equation}}
\def\bea{\begin{eqnarray}}
\def\eea{\end{eqnarray}}

\newcommand{\jfi}
{\affiliation{James Franck Institute, The University of Chicago, Chicago, IL 60637}}

\newcommand{\gaug}
{\affiliation{Institute for Theoretical Physics, Georg-August-Universit\"at G\"ottingen, 37077 G\"ottingen, Germany}}

\begin{document}
\setstcolor{red}
\title{Robustness of travelling states in generic non-reciprocal mixtures}

\author{Rituparno Mandal}%
\email[Email: ]{rituparno.mandal@uni-goettingen.de}
\gaug
\jfi

\author{Santiago Salazar Jaramillo}
\gaug

\author{Peter Sollich}%
\email[Email: ]{peter.sollich@uni-goettingen.de}
\gaug
\affiliation{Department of Mathematics, King's College London, London WC2R 2LS, UK}

\begin{abstract}

Emergent non-reciprocal interactions violating Newton's third law are widespread in out-of-equilibrium systems. Phase separating mixtures with such interactions exhibit travelling states with no equilibrium counterpart. Using extensive Brownian dynamics simulations, we investigate the existence and stability of such travelling states in a generic non-reciprocal particle system. By varying a broad range of parameters including aggregate state of mixture components, diffusivity, degree of non-reciprocity, effective spatial dimension and density, we determine that travelling states do exist below the predator-prey regime, but nonetheless are only found in a narrow region of the parameter space. Our work also sheds light on the physical mechanisms for the disappearance of travelling states when relevant parameters are being varied, and has implications for a range of non-equilibrium systems including non-reciprocal phase separating mixtures, non-equilibrium pattern formation and predator-prey models.

\end{abstract}

\maketitle

\section{Introduction}

Fundamental pairwise interactions such as gravitational or electromagnetic forces always obey action-reaction symmetry. This is also true for effective pairwise interactions between particles in an equilibrium medium like the Asakura–Oosawa attraction~\cite{asakura1954} between colloidal particles in a medium of non-adsorbing macromolecules. Similar examples arise in the context of Casimir forces between compact objects~\cite{kardar2007}. This paradigm can break down if either the medium or the interacting particles are driven out of equilibrium~\cite{lowen2015}, resulting in non-reciprocal interactions~\cite{mandal2022,sriram2022}; these have generated very substantial interest in the last decade~\cite{lowen2003,bartnick2015,lisin2017, kryuchkov2018,lin2018,loos2019,kryuchkov2020,dadhichi2020,falcao2020, golestanian2020,marchetti2020,loos2020,vitelli2021,carletti2022,zhang2022, packard2022,golestanian2019,kreienkamp2022}.

A striking observation is that non-reciprocal systems can give rise to exotic time-dependent steady states~\cite{vitelli2021, marchetti2020, golestanian2019, golestanian2020} with stable travelling waves. Although hydrodynamic continuum descriptions~\cite{vitelli2021,marchetti2020,golestanian2020,hulsmann2021,hulsmann2021_localized,brauns2023,hulsmann2023,alston2023,suchanek2023} have shed some light on the conditions under which travelling states arise and their stability with respect to changes in material parameters, similar insights from particle-based models remain scarce. One instance is a study of particles with long-ranged non-reciprocal
diffusiophoretic interactions mediated by concentration gradients in the surrounding medium~\cite{golestanian2019} and another involves quorum-sensing in active matter \cite{duan2023}.

\begin{figure}[h]
\includegraphics[width =0.99\linewidth]{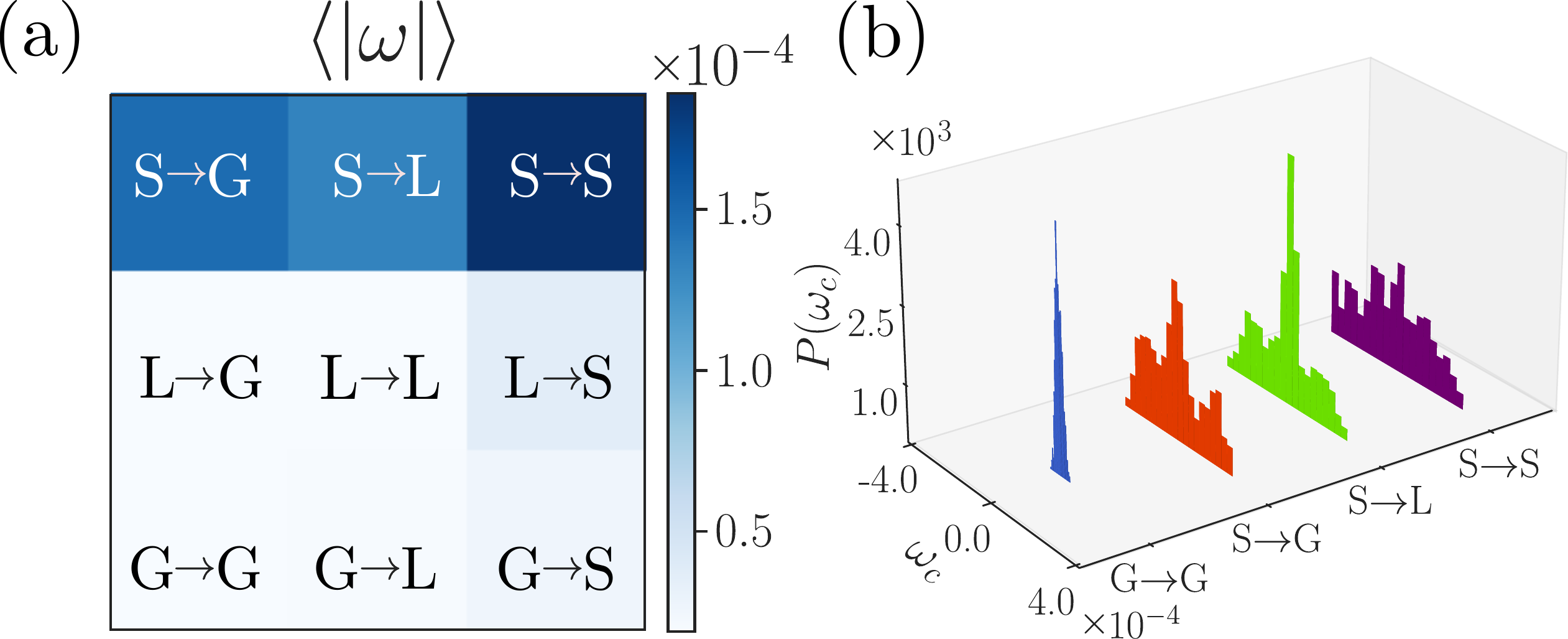}
\caption{(a) Mean angular velocity in annular geometry (with $R_{\rm in}=33$ and $R_{\rm out}=42$) for all combinations of aggregate states {\bf S}olid, {\bf L}iquid and {\bf G}as \textit{e.g.} for A in solid and B in gas state we use $V_{\rm AA}=1$, $V_{\rm BB}=0$; see main text and SM~\cite{SM} for details). The arrow indicates which species has the higher attraction towards the other species and will therefore tend to follow it as the ``chasing cluster'' in a travelling state.We use density $\rho=0.1$, diffusivity $D_0=10^{-2}$ and $\delta = \pm 0.9$. (b) Probability distribution of centre of mass angular velocity $\omega_c$, for four combinations of aggregate states as indicated. Note the clearly visible peaks at nonzero $\omega_c$ for S $\to$ G. The distribution for G $\to$ G has been scaled down by a factor of 4 for better visualisation.}
\label{fig:state}
\end{figure}

Here we explore a generic particle-based model of a non-reciprocal system with short-ranged interactions, employing Brownian dynamics simulations to investigate the existence and stability of collective travelling states. The particle-based approach implements non-reciprocal interactions directly at the microscopic level, unlike in hydrodynamic models where non-reciprocal effects have to be included via coarse-graining which, with rare exceptions~\cite{sollich2022}, cannot be done exactly. We systematically vary the most relevant parameters that potentially influence travelling states, such as the diffusivity $D_0$ and density of the constituent particles $\rho$, the aggregate state (gas, liquid, solid) of the mixture components, the degree of non-reciprocity $\delta$,  the degree of confinement and with it the effective spatial dimension. We identify the physical mechanisms that maintain or destroy travelling states, and find that the requirements for such states are met only in a narrow region of the parameter space.

\begin{figure*}
\centering
\includegraphics[height =.45\linewidth]{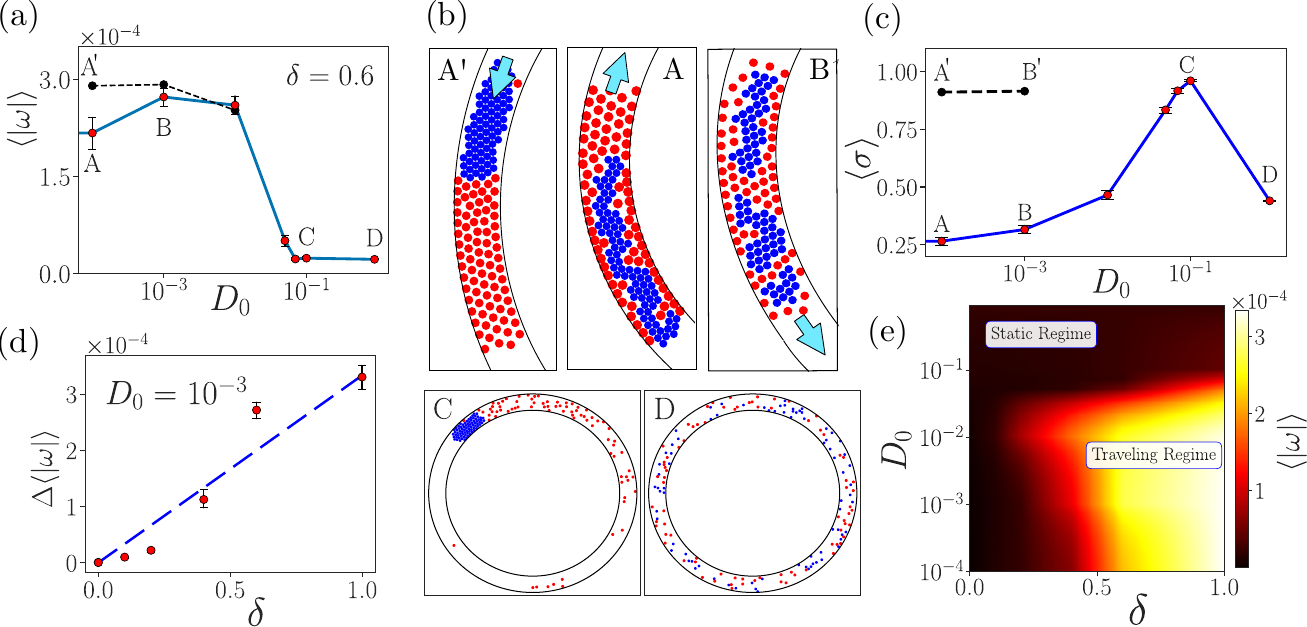}
\caption{Effects of diffusivity and non-reciprocity, for a system with the same density and annular geometry as in Fig.~\ref{fig:state}. (a) Average absolute angular velocity $\langle |\omega|\rangle$ of the particles plotted against diffusivity $D_0$, for a system with $\delta=0.6$ initialized in a phase separated state (dashed line and black filled circles) and for a mixed initial condition (solid blue line with red filled circles). (b) Snapshots of the binary mixture (direction of motion in the travelling cases is marked by arrows) for different $D_0$ and initial conditions as marked in (a). (c) Phase segregation order parameter $\langle \sigma \rangle$~\cite{SM} as a function of $D_0$ for mixed and segregated initial conditions (color scheme and $\delta=0.6$ as in (a)). (d) Dependence of net mean angular velocity $\Delta \langle |\omega|\rangle$ (see text for definition) on non-reciprocity parameter $\delta$ for $D_0=10^{-3}$. (e) Phase diagram from heat map of  $\langle|\omega|\rangle$ against diffusivity $D_0$ and non-reciprocity parameter $\delta$: travelling states appear when non-reciprocity is high {\em and} 
diffusivity is sufficiently low.}
\label{fig:phase}
\end{figure*}

\begin{figure}
\includegraphics[width = 0.99\linewidth]{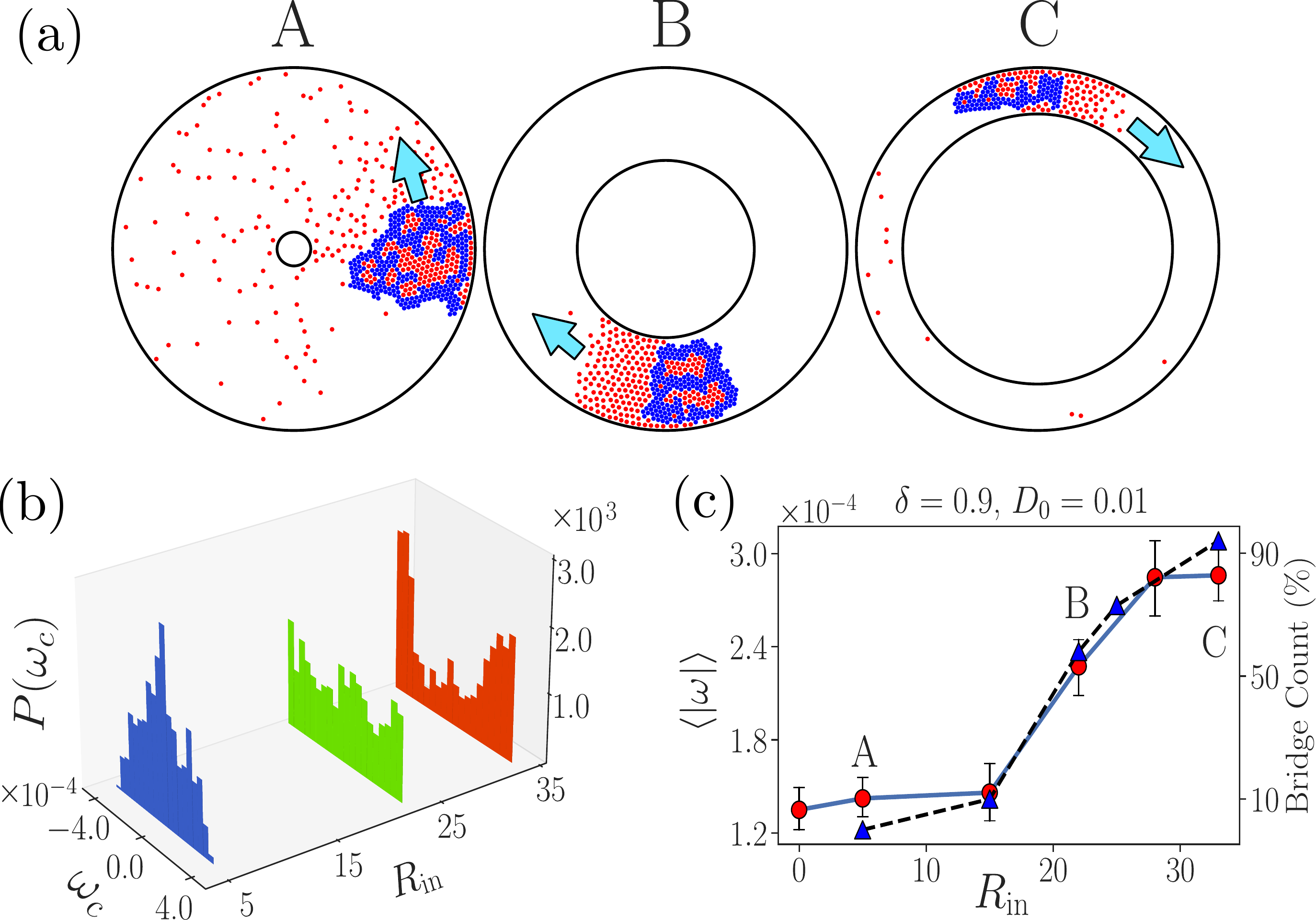}
\caption{(a) Typical snapshots of the non-reciprocal binary mixture in an annular geometry (same density $\rho=0.1$ and outer radius $R_{\rm out}=42$ as in Fig.~\ref{fig:state}, $\delta=0.9$, $D_0=0.01$) for 
inner radius $R_{\rm in}=5$ (A), $22$ (B) and $33$ (C), respectively. (b) Distribution for the angular velocity of the center of mass $\omega_c$ for three different $R_{\rm{in}}$, showing clear peaks in the narrow channel regime ($R_{\rm in}=33$). (c) Red dots: Average absolute angular velocity $\langle |\omega|\rangle$ as a function of the inner channel radius $R_{\rm{in}}$. Blue triangles: Percentage of bridge-like structures, which follows an almost identical trend.} 
\label{fig:dimension}
\end{figure}

\paragraph{Model:}
We study a binary mixture of species A and B (See Supplemental Material~\cite{SM}, for details on the model and numerical solution)
in two dimensions with Lennard-Jones (LJ) interactions, with homogeneously mixed initial conditions unless stated otherwise. The  intra-species (AA or BB) interaction is reciprocal; we tune the strength of its attractive part to make the individual components of the mixture (A or B), in pure form and at low temperature equilibrium, behave as a gas, liquid or solid. When phase separation occurs in the combined (50\% A, 50\% B) system, we find that the A- and B-rich phases then also have gas-, liquid- and solid-like features, and 
label them accordingly. For the inter-species (AB, BA) interaction we take the repulsive part ($\sim 1/r^{12}$) as reciprocal, thus defining the physical core size of the particles. The attractive part ($\sim 1/r^{6}$) is non-reciprocal, with a prefactor $1\pm \delta$ where $\delta$ is the non-reciprocity parameter. For $\delta>0$ there exists a stronger attractive force on A-particles (colored blue in plots) from neighbouring B-particles (colored red) than on Bs from surrounding As. 
We focus on the range $0<\delta<1$ but have also explored the predator-prey regime of larger $\delta$, {\it{e.g.}} $\delta \sim 1.5$, finding similar conclusions. We set the overall strength of the AB and BA attraction such that phase separation between A- and B-rich phases can occur, which is a precondition for travelling states.

We implement both periodic boundary conditions (PBC~\cite{SM}, 
with square box of linear size $L=42$) and confined annular geometries (see SM ~\cite{SM} for the advantages of this choice). To explore whether analogs of the travelling states found in hydrodynamic models~\cite{vitelli2021, marchetti2020, golestanian2020} exist in our particle-based setup, we measure the average rotational velocity $\langle |\omega| \rangle$ (for annular geometry) or average translational velocity $\langle |\bf{v}| \rangle$ (in PBC or effectively 2$d$ geometries), with the average $\langle \ldots\rangle$ being taken across all particles. We also analyse $P(\omega_c)$ or $P(v_c)$ (where $\omega_c$ and $v_c$ are the angular and translational velocity of the centre of mass) as a diagnostic: a probability peak 
at nonzero $\omega_c$ or $v_c$ indicates a travelling state in the annular geometry and PBC, respectively. The samples for these 
distributions are collected both across time in steady state and the ensemble of trajectories~\cite{SM}. 

\paragraph{Effects of aggregate state of mixture components:}
Physically, the aggregate state of each mixture component should be key in the emergence of travelling states. We therefore tune the intra-species attraction strengths $V_{\rm AA}$ and $V_{\rm BB}$ such that the pure species A and B at equilibrium at low diffusivity (see below) are in a gas, liquid or solid state~\cite{SM}. In Fig.~\ref{fig:state}(a) we show the behaviour for the nine resulting combinations 
for an annular geometry%
; for typical configurations see Fig.~\ref{fig:phase}(b) and Fig.~S1~\cite{SM}. 
A travelling state where the chasing cluster achieves a reasonable velocity appears only when the chasing particles have strong inter-particle  attractions. The same conclusion turns out to hold for other boundary conditions, e.g.\ PBC (see Fig.~S2~\cite{SM}). The distribution of the center of mass angular velocity is generally broad when the chasing particles are in a solid state. Characteristic velocity peaks appear only when a solid cluster is chasing a gas (Fig.~\ref{fig:state}(b)), making this combination the most promising candidate for travelling states. This observation also holds in the predator-prey regime $\delta>1$ and for larger system sizes (see 
Fig.~S8 and Fig.~S10~\cite{SM}). We therefore adopt the corresponding interaction strengths ($V_{\rm AA}=1$, $V_{\rm BB}=0$) for the rest of the analysis.
 
\paragraph{Effects of diffusivity and degree of non-reciprocity:}
Diffusivity, which plays the role of thermal fluctuations provided by the embedding medium, is another crucial factor in the collective dynamics. Fig.~\ref{fig:phase}(a) shows that the mean angular velocity of a non-reciprocal  mixture generally decreases with increasing diffusivity $D_0$%
. To understand this, one has to bear in mind that the actual aggregate state of the mixture components varies with changing diffusivity. Indeed, the snapshots in Fig.~\ref{fig:phase}(b) demonstrate that at high $D_0$ (point D), both mixture components are in a gas state and not segregated, both of which prevent a travelling state. Interestingly, the system in Fig.~\ref{fig:phase}(b) at point C does possess one dense and one gaseous phase and is completely segregated between A and B, as shown by a high segregation order parameter $\langle \sigma \rangle$~\cite{SM} in Fig.~\ref{fig:phase}(c). Nonetheless, persistently high fluctuations in the distribution of A-particles prevent a travelling state also here: we conclude that segregation into a dense and gaseous phase, even alongside a large non-reciprocity parameter $\delta$, does not by itself guarantee the existence of a travelling phase.

We next study the behaviour for moderate $D_0$. Here Fig.~\ref{fig:phase}(a, blue line) shows an initially surprising intermediate maximum in the mean angular velocity. This is caused by incomplete separation of the mixture within our finite simulation time (see Fig.~S3~\cite{SM}) at low $D_0$ when the system is initialized in a mixed state. A system starting in a completely segregated state (dashed line in Fig.~\ref{fig:phase}(a)), on the other hand, has an angular velocity decreasing monotonically with $D_0$. Consistently with this picture, the degree of segregation $\langle \sigma \rangle$ of the mixture from a mixed initial state is first facilitated kinetically by increasing $D_0$ (Fig.~\ref{fig:phase}(c)), until it reaches a maximum. Afterwards, it decreases as both components become gaseous and mix.
When initializing in a segregated state, on the other hand (Fig.~\ref{fig:phase}(c), dashed line), there is no issue in reaching a segregated state kinetically and the degree of segregation decreases essentially monotonically with $D_0$.

The dependence of the mean angular velocity on the non-reciprocity parameter $\delta$ is rather simpler: monotonically increasing and approximately linear when diffusivity is kept constant ($D_0=10^{-3})$. Fig.~\ref{fig:phase}(d) shows $\Delta \langle |\omega|\rangle \equiv \langle |\omega|\rangle -\langle |\omega|\rangle_0$; the second term subtracts off the uninteresting fluctuations of $\omega$ in the reciprocal case $\delta=0$. The linearity suggests that the interface between the A and B-phases remains largely unaffected by increasing $\delta$, with just the interfacial driving force increasing linearly as the imbalance between AB and BA-attractions is proportional to $\delta$.

A phase diagram can be created from the above results by plotting the mean angular velocity as a function of $\delta$ and $D_0$. In Fig.~\ref{fig:phase}(e) lighter colors correspond to higher $\langle |\omega|\rangle$, {\em i.e.}\ pronounced travelling states (cf.\ similar results for PBC~\cite{SM}, Figs.~S4, S5). The appearance of a travelling regime at high non-reciprocity and low diffusivity is also found in the predator-prey regime (Fig.~S9 \cite{SM}) and for larger system sizes  (Fig.~S11 \cite{SM}), and is consistent with earlier studies~\cite{bartnick2016,marchetti2020, golestanian2020}.
 
\paragraph{Effect of geometry and dimensionality:} Next we study the behaviour of travelling states as a function of the geometry and hence (effective) spatial dimension (while keeping 
density, non-reciprocity diffusivity
fixed). In the annular geometry, we confine our system between two concentric circles of radius $R_{\rm in}$ and $R_{\rm out}$. For $R_{\rm in} \simeq R_{\rm out}$ the system is, up to small corrections from the annulus curvature~\cite{SM}, 
equivalent to a straight channel (with PBC in the direction of its long axis) 
as considered in~\cite{marchetti2020},
whereas for $R_{\rm in}=0$ we have a confined 2$d$ system. We observe travelling states (see Fig.~\ref{fig:dimension}(a) for snapshots) to be more persistent in the narrow channel limit while with decreasing $R_{\rm in}$ the motion becomes more erratic (see Fig.~S6 for persistence time data and Fig.~S7 for the distribution of the linear velocity magnitude%
~\cite{SM}). 
The distribution of the center of mass angular velocity $P(\omega_c)$ shows characteristic peaks at nonzero $\omega_c$ (and its negative) at larger values of $R_{\rm in}$, while it is unimodal for $R_{\rm in}=5$ (Fig.~\ref{fig:dimension}(b)). The mean angular velocity also decreases with the inner radius (Fig.~\ref{fig:dimension}(c)), indicating that travelling phases become more transient as the effective dimension changes from one to two.

Inspection of Fig.~\ref{fig:dimension}(a) suggests that this behaviour of travelling states at smaller $R_{\rm in}$ sets in when the travelling solid cluster no longer forms a bridge touching both circular walls. Particles from the gas phase can then ``leak'' past the cluster (see supplementary movie~\cite{SM}), causing it to reverse direction. To substantiate this hypothesis we measured the probability for configurations to contain bridges. This bridging probability decreases significantly with the inner radius as indicated by Fig.~\ref{fig:dimension}(c) and in fact follows an almost identical trend to $\langle |\omega| \rangle$. It should be noted that our 2$d$ ($R_{\rm in}=0$) system is confined by a physical wall, which makes it distinct from systems with PBC where clusters can bridge across periodic boundaries~\cite{golestanian2020}.

\begin{figure}
\includegraphics[width =0.99\linewidth]{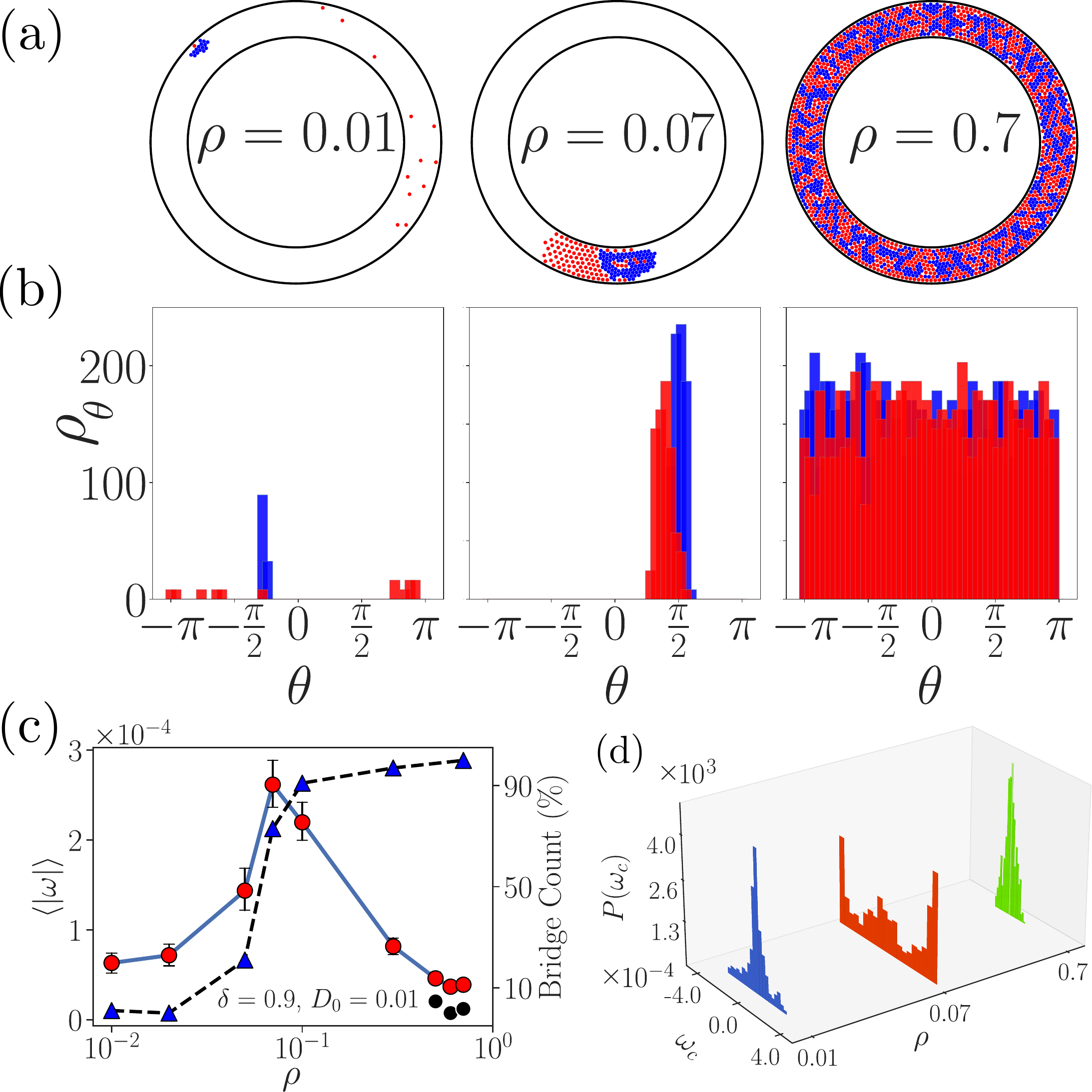}
\caption{(a) Snapshots of systems with densities $\rho=0.01$, $0.07$, $0.7$ (other parameters as 
in Fig.~\ref{fig:state}: $D_0=0.01$, $\delta=0.9$, $R_{\rm in}=33$, $R_{\rm out}=42$
). (b) Angular distribution of particle number density for the configurations in (a). (c) Red dots: average angular velocity $\langle|\omega|\rangle$ versus density. Black dots: results for a completely segregated, rather than mixed, initial condition. Blue triangles: probability of bridge formation. (d) Distribution of the center of mass angular velocity $\omega_c$, with 
clear peak demonstrating a travelling state at intermediate $\rho$.}
\label{fig:density}
\end{figure}

\paragraph{Effects of density:} Finally, we study how the overall number density $\rho$ 
of a non-reciprocal mixture influences the existence of travelling states; Fig.~\ref{fig:density}(a) shows snapshots at three exemplary $\rho$,
Fig.~\ref{fig:density}(b) 
the instantaneous angular density distributions. The mean angular velocity $\langle |\omega|\rangle$ exhibits a non-monotonic response with density (Fig.~\ref{fig:density}(c)). This can be rationalised from the fact that a well-defined asymmetric interface between two species only forms at an intermediate density. Indeed, Fig.~\ref{fig:density}(b) indicates that at low $\rho$, A and B particles are typically far apart. For moderate $\rho$, further analysis reveals that the small angular velocities  are correlated with the bridging probability as before (Fig.~\ref{fig:density}(c), blue triangles): only once a bridge between inner and outer walls exists, will a gas of B particles ``pile up'' on one side of a travelling A-cluster and so form an AB-interface there. This behaviour is not due to any symmetry breaking between the outer and inner radius, since the correlation between the appearance of bridges and the velocity of the traveling state is also found in a quasi-1D system, cf.~Fig.~S12 in the SM~\cite{SM}.

For different reasons, the system also loses the asymmetric interface between the two species at high $\rho$. Starting from a mixed initial state, the high density hinders the segregation kinetics so that A and B particles only form small domains without macroscopic interfaces (see snapshot and angular density for $\rho=0.7$ in Fig.~\ref{fig:density}(a,b)). Even starting from segregated initial conditions (black points in Fig.~\ref{fig:density}(c)), however, no travelling state occurs because there is not enough space to create the necessary density gradient of the gaseous B species. With a nearly constant density within the B-phase, the two macroscopic interfaces between A and B are then symmetric with each other so that the net unbalanced forces generated at each individual interface cancel. As a result, intermediate densities are optimal for generating travelling states, where the probability distribution of the center of mass angular velocity  (Fig.~\ref{fig:density}(d)) again shows clear peaks at nonzero $\omega_c$.

\begin{figure}
\includegraphics[height =.43\linewidth]{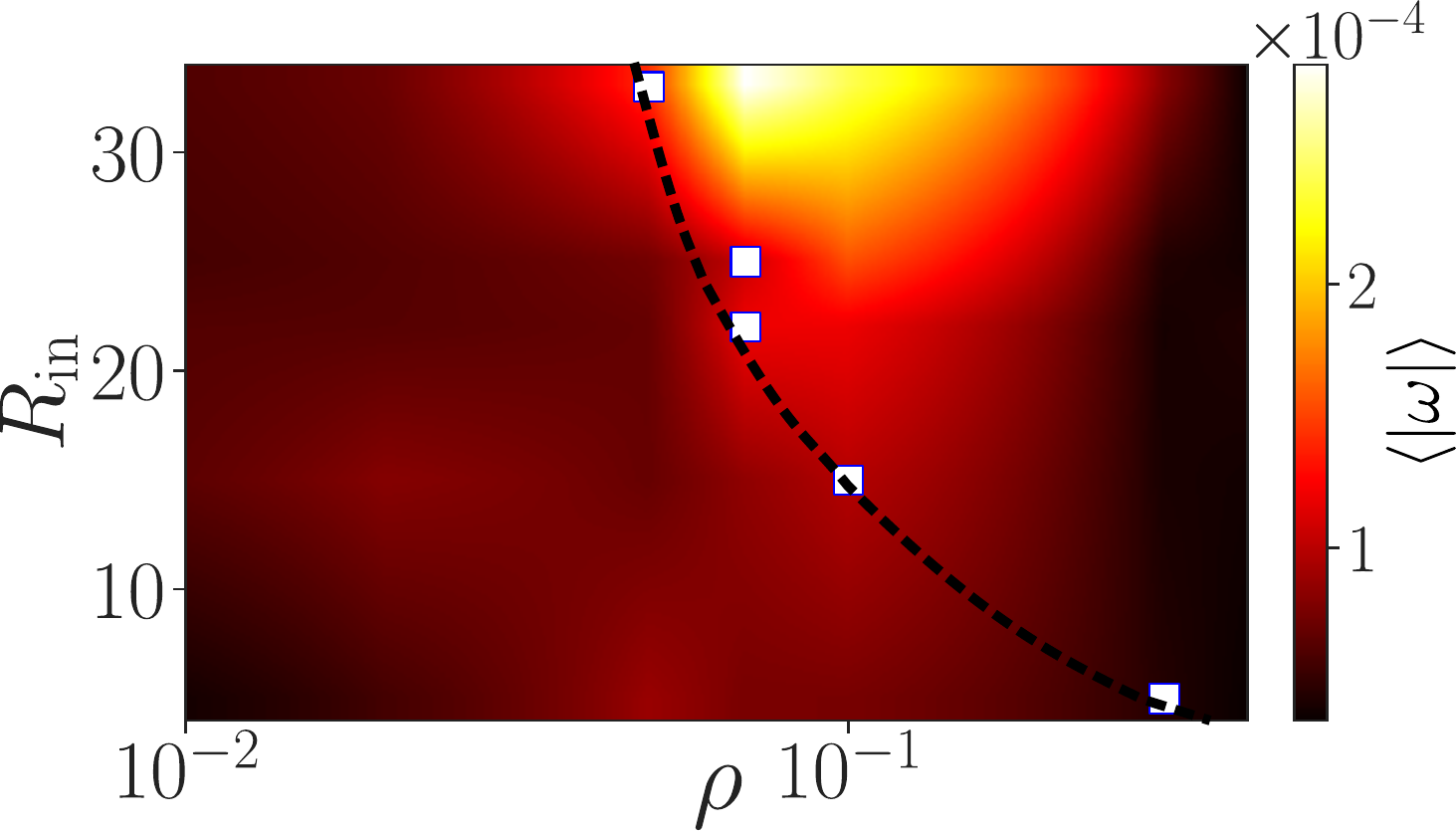}
\caption{Phase diagram from heat map of $\langle | \omega | \rangle$ as a function of density $\rho$ and inner radius with $\delta = 0.9$, $D_0=0.01$ and $R_{\rm{out}}=42$. White squares: densities $\rho$  where the bridge probability surpasses $20\%$ for each radius. Dashed line: guide to the eye.}
\label{fig:bridge}
\end{figure}

Finally we combine the data from variation of the inner radius and  the density into a second phase diagram (see Fig.~\ref{fig:bridge}). The white squares have been obtained by setting a threshold on the bridging probability and are seen to demarcate travelling and non-travelling states well in the low density part of the phase diagram. In the high-density region, the difficulty of forming asymmetric interfaces kicks in to prevent the formation of travelling states. 

Summarizing, we have studied the emergence and robustness of travelling states in a non-reciprocal binary particle mixture. We find that such states, with an appreciable velocity, appear in the low non-reciprocity regime only when a ``chasing'' cluster with a solid-like structure is present. The travelling velocity increases with increasing degree of non-reciprocity and with {\em decreasing} diffusivity, because of stronger segregation between the phases. Varying the system geometry, travelling states become more persistent as we move from a (confined) two-dimensional scenario to an effectively one-dimensional annulus, where the solid cluster can form a bridge between inner and outer walls. 
This effect also facilitates travelling states at intermediate density and hinders them for dilute systems; high density also suppresses travelling states because asymmetric interfaces can no longer form. Thus, travelling states may only be found in a narrow region of the parameter space, where the conditions for their appearance are optimal. We have also found that most of our conclusions hold for the predator-prey regime ($\delta>1$) and for increasingly larger system sizes (as detailed in the SM~\cite{SM}), although further studies are required in order to ascertain the validity of our conclusions in the thermodynamic limit.

Our results should be amenable to verification in experiments on non-reciprocal particle  systems~\cite{speck2018,meredith2020}. Our identification of the physical mechanisms generating travelling states will also be of broader importance for understanding the behaviour of non-reciprocal phase separating mixtures and non-equilibrium pattern formation more generally, including -- outside of physics -- in e.g.\ prey-predator models.

\paragraph*{Acknowledgements:}
RM and SSJ contributed equally to this work. This project has received funding from the European Union’s Horizon 2020 research and innovation programme under the Marie Sk\l odowska-Curie grant agreement No 893128. Simulations were run on the GoeGrid cluster at the University of Göttingen, which is supported by the Deutsche Forschungsgemeinschaft (project IDs 436382789; 493420525).

\bibliography{nonreciprocal}
\bibliographystyle{ieeetr}

\end{document}


\title{Supplementary material for: Robustness of travelling states in generic non-reciprocal mixtures}

\author{Rituparno Mandal}%
\email[Email: ]{rituparno.mandal@uni-goettingen.de}
\gaug

\author{Santiago Salazar Jaramillo}
\email[Email: ]{s.salazarjaramillo@uni-goettingen.de}
\gaug

\author{Peter Sollich}%
\email[Email: ]{peter.sollich@uni-goettingen.de}
\gaug
\affiliation{Department of Mathematics, King's College London, London WC2R 2LS, UK}

\maketitle

\newpage

\section{Model}
\label{sec:model}

 To simulate a generic non-reciprocal particle based system we use a binary mixture of soft particles interacting via a modified Lennard-Jones potential
\begin{equation}\label{eq:LJ_potential}
V_{\alpha\beta}(r_{ij}) = 4\epsilon_0 \left( \Big(\frac{\sigma_0}{r_{ij}}\Big)^{12} - D_{\alpha\beta} \Big(\frac{\sigma_0}{r_{ij}}\Big)^{6} \right)
\end{equation}
where the relative distance is $r_{ij}=|{\bf{r}}_i-{\bf{r}}_j|$ and ${\bf{r}}_i$, ${\bf{r}}_j$ are the position vectors of particles $i$ and $j$. The Greek indices $\alpha, \beta$ refer to the particle type (A or B) of particles $i$ and $j$, respectively. The energy scale is set by $\epsilon_0=1.0$, the length scale is set by $\sigma_0=1.0$ and the time scale is given in units of $\gamma_0^{-1}$, which is explained in the following. We also truncate the potential at $r_c=2.5 \sigma_0$ to make the interaction short ranged for computational benefits. We define $D_{\alpha \beta}$ as the elements of the non-reciprocity matrix 
\begin{equation}
\label{eq:nonrec_matrix}
 D = \begin{pmatrix}
V_{\rm AA} & \overline{V}_{\rm AB}(1+\delta)  \\
\overline{V}_{\rm AB}(1-\delta) & V_{\rm BB}   \end{pmatrix} 
\end{equation}
such that the diagonal elements represent the strength of the attractive interaction between particles of the same type. The intra-species interactions are reciprocal by construction. We tune $V_{\rm AA}$ and $V_{\rm BB}$ to determine the equilibrium aggregate state of the pure A and B species at low diffusivity (see below). For $V_{\rm AA}=1$ we observe a solid state with an ordered, practically rigid particle cluster. For $V_{\rm AA}=0$, on the other hand, the like-particle interaction is completely repulsive and the system behaves as a gas. For $V_{\rm AA}=0.5$ we see clustering but without prominent crystalline order and so we use this parameter value to obtain a liquid state. We checked the radial distribution functions of the equilibrium configurations generated by these parameter settings, to make sure that they correspond to the intended phases of gas, liquid and solid.

As $D_{\alpha \beta}$ is asymmetric  in general, we need to specify that Eq.~(\ref{eq:LJ_potential}) represents the interaction potential that particle $i$ (of type $\alpha$) feels due to particle $j$ (of type $\beta$). In the off-diagonal terms, $\overline{V}_{\rm AB}$ and $\delta$ tune the strength of the inter-species interactions and the degree of non-reciprocity, respectively.
Note that we are tuning the non-reciprocity only via the attractive part of the interaction, which we argue is physically reasonable as the repulsive term sets a (soft) core size of the particles.
For all the results mentioned in this article we have kept $\overline{V}_{\rm AB}=0.25$ and varied $\delta$ to control the degree of non-reciprocity in the model.

To simulate the dynamics, we numerically solve the Langevin equation in two spatial dimensions using the Euler-Maruyama method \cite{lennon2008}, i.e.\ the update rule
\begin{equation}
    \mathbf{r}_i(t+\Delta t) = \mathbf{r}_i(t) - \frac{1}{\gamma_0} \nabla_i \sum_{j }V_{\alpha\beta}(r_{ij})\;\Delta t + \sqrt{2 D_0 \Delta t}\;\; \boldsymbol{\xi}_i\,.
    \label{eq:discrete_langevin}
\end{equation}
where $\gamma_0$ is the coefficient of friction and $\Delta t$ is the integration time step. Here $D_0$ is the diffusion constant and $\boldsymbol{\xi}_i$ is the zero mean unit variance Gaussian white noise acting on particle $i$. 

To implement the boundary condition we use a steep confining potential that increases harmonically as particles leave the region between two concentric circles of radii $R_{\rm in}$ and $R_{\rm out}$, generating an annulus geometry.
As an alternative we consider a square simulation box (with a linear size $42$) with periodic boundaries. Most of our results are obtained for the annular geometry as it is an experimentally realizable setup and very useful to explore non-equilibrium travelling states~\cite{denis2019,buhl2006, souslov2017, bricard2013,pearce2015,joshi2023, kolpas2007}. Depending on overall density we set the number of particles $N$ to lie between 30 and 1800 and use $\Delta t= 10^{-4}$, running simulations for $10^8-10^9$ time steps. Steady state time averages are obtained from the second half of the total run time, with the first half being discarded as burn-in unless stated otherwise. To improve statistical accuracy we use an ensemble of $\sim 200$ independent simulations for each parameter setting. A similar sampling scheme is implemented when collecting samples for the probability distributions, the first half of the simulations is discarded and the remaining data is sampled in time (collected over significant time interval $\Delta \tau=10^3-10^4$ depending on the length of the run, to avoid correlations) for each configuration of the ensemble consisting of $\sim 200$ independent trajectories.

\begin{figure}[ht]
\includegraphics[height =0.95\linewidth]{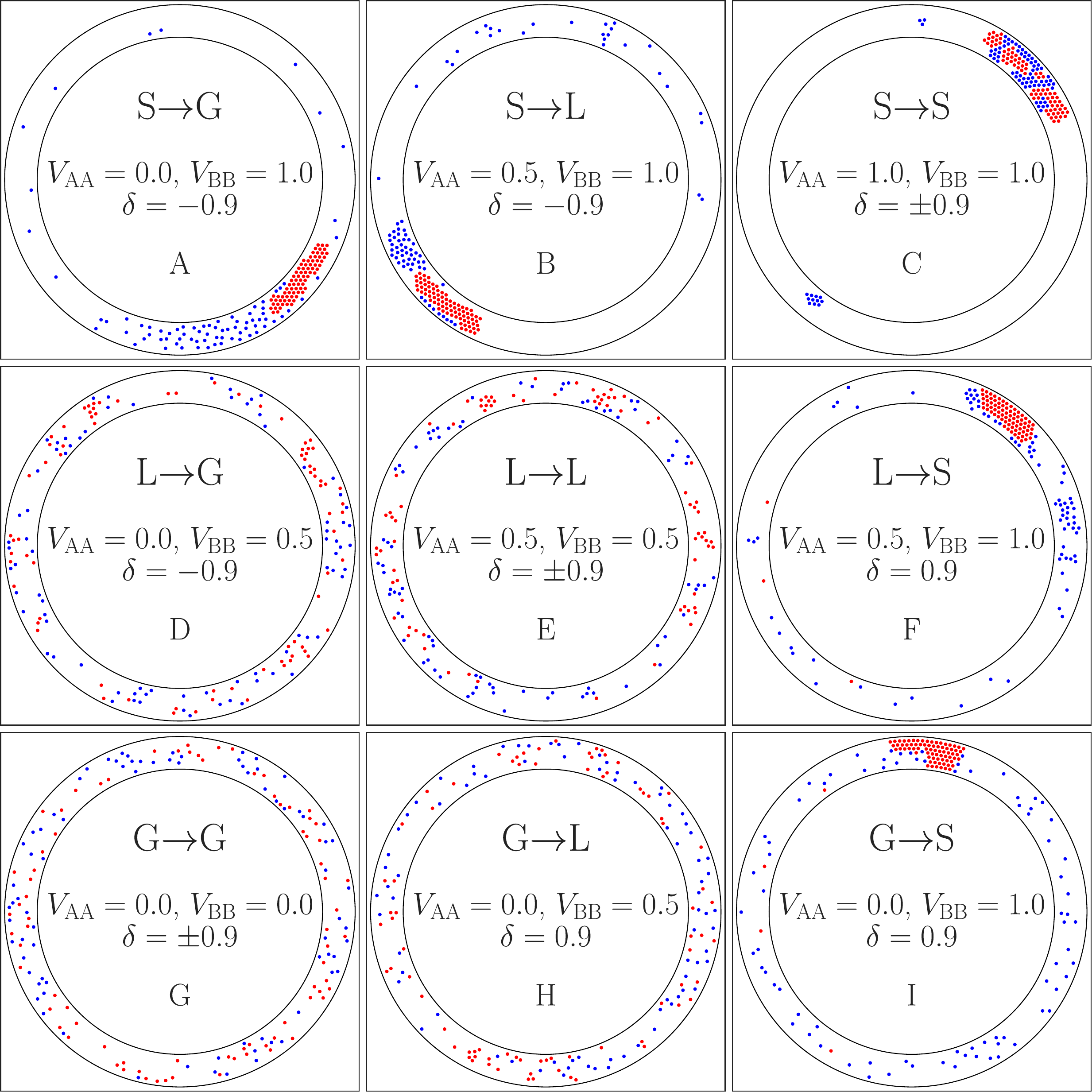}
\caption{Snapshots of the system for the annular geometry shown in Fig.~1 of the main text. A solid state is characterized by a dense cluster of particles with a rigid body-like motion, while in the gas states the particles move largely independently of each other. The liquid state exhibits a degree of cohesion but lacks crystalline order. The sign of the non-reciprocity parameter controls the direction of the asymmetry in the interaction, changing which species ``chases'' and which species ``runs''.} 
\label{fig:snap_state}
\end{figure}

\begin{figure}[ht]
\includegraphics[height=0.7\linewidth]{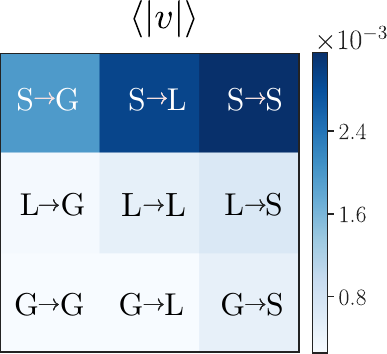}
\caption{Mean translational velocity for $\delta = \pm 0.9$ for different values of the interaction parameters (same set of parameters as in Fig.~1 of the main text) in a two dimensional square box with periodic boundary conditions.}
\label{fig:multi_state_pbc}
\end{figure}

\begin{figure}[ht]
\includegraphics[height=0.75\linewidth]{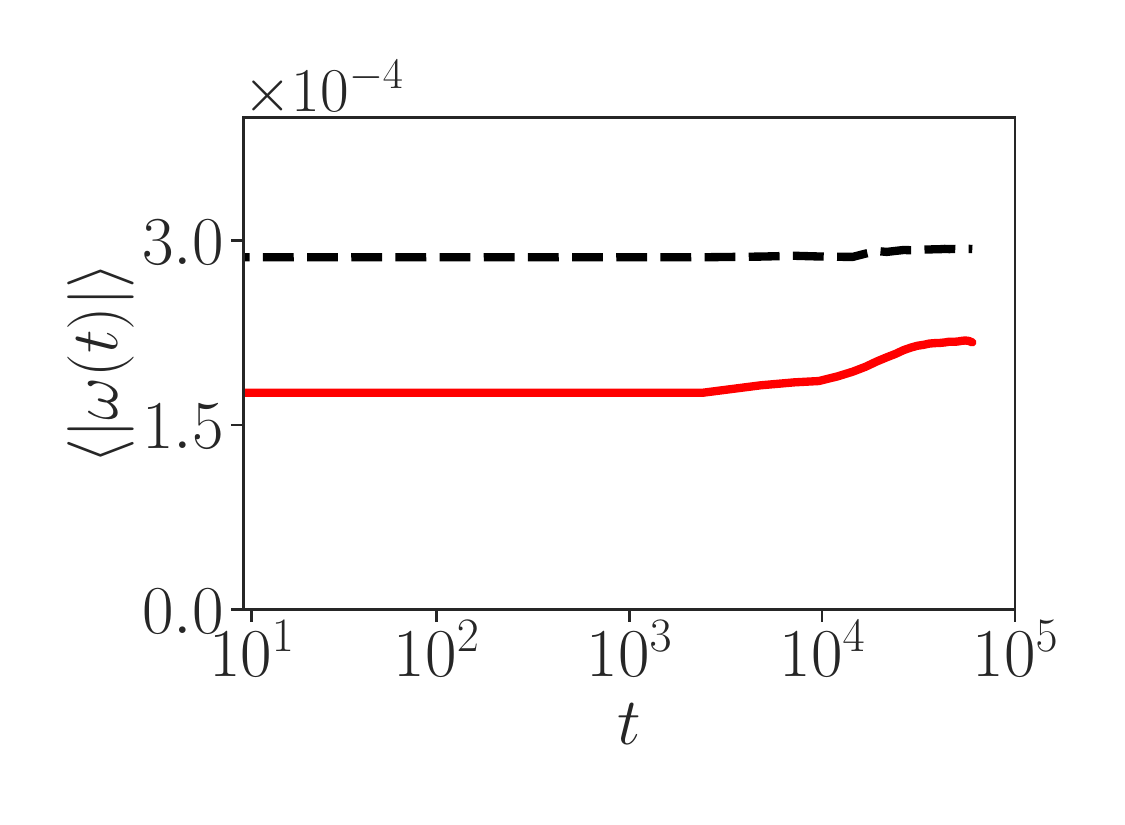}
\caption{Time evolution of the mean angular velocity for a system with $\delta = 0.6$ and $D_0= 10^{-4}$. The red line shows the evolution of a system initialized in a random initial configuration, corresponding to the point marked as A in Fig.~2(a) from the main text. The dashed lines represents point $\rm{A}^\prime$ in the same figure, a system with a segregated initial configuration.}
\label{fig:time_evol}
\end{figure}

\begin{figure}[ht]
\includegraphics[height =0.5\linewidth]{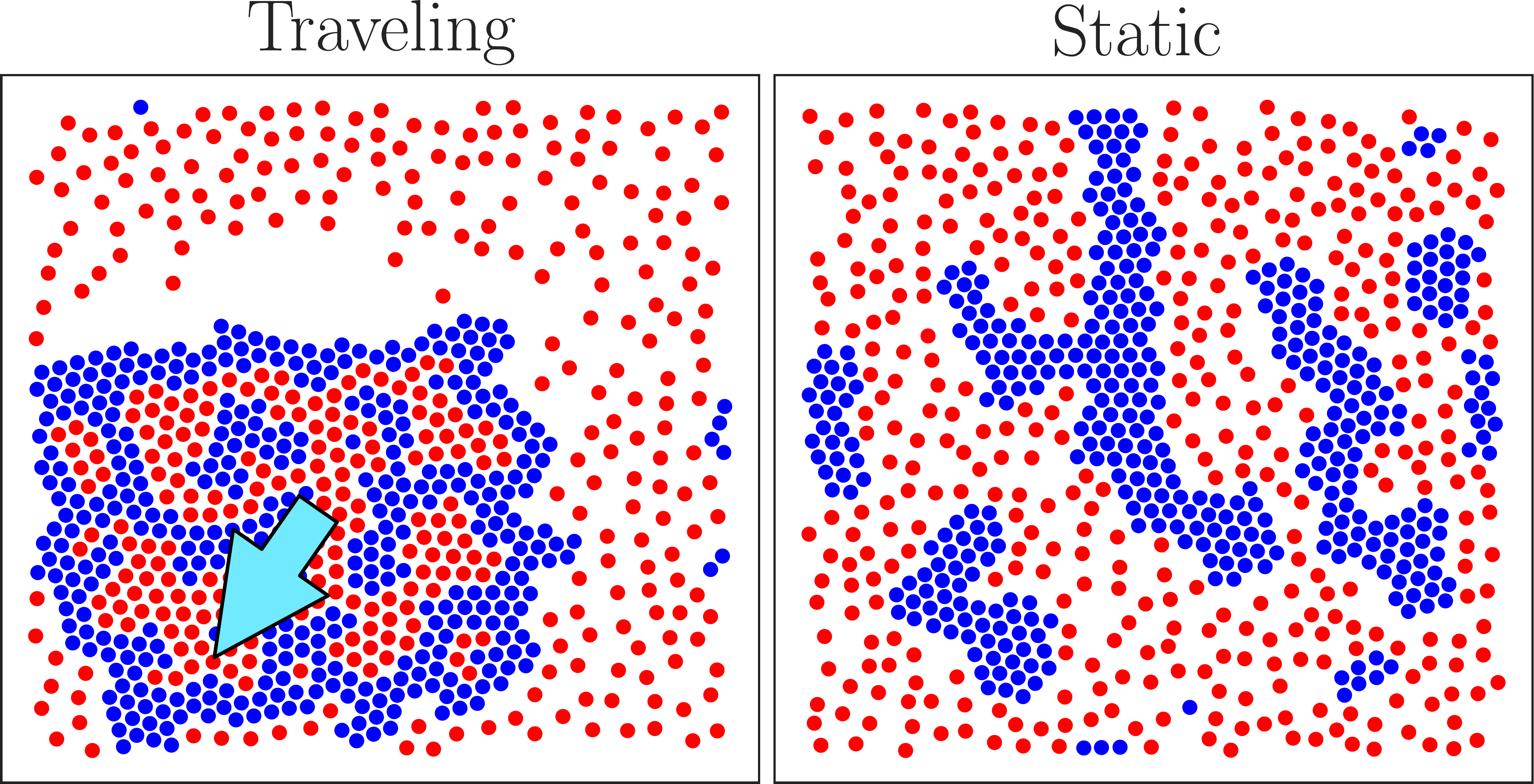}
\caption{Snapshots of systems with periodic boundary conditions for parameters chosen from the phase diagram shown in Fig.~2(e) of the main text. The travelling state corresponds to a system with $\delta = 0.9$, $D_0 = 0.01$ and the static one to $\delta = 0.4$, $D_0 = 0.07$.}
\label{fig:state_pbc}
\end{figure}

\begin{figure}[ht]
\includegraphics[height =0.62\linewidth]{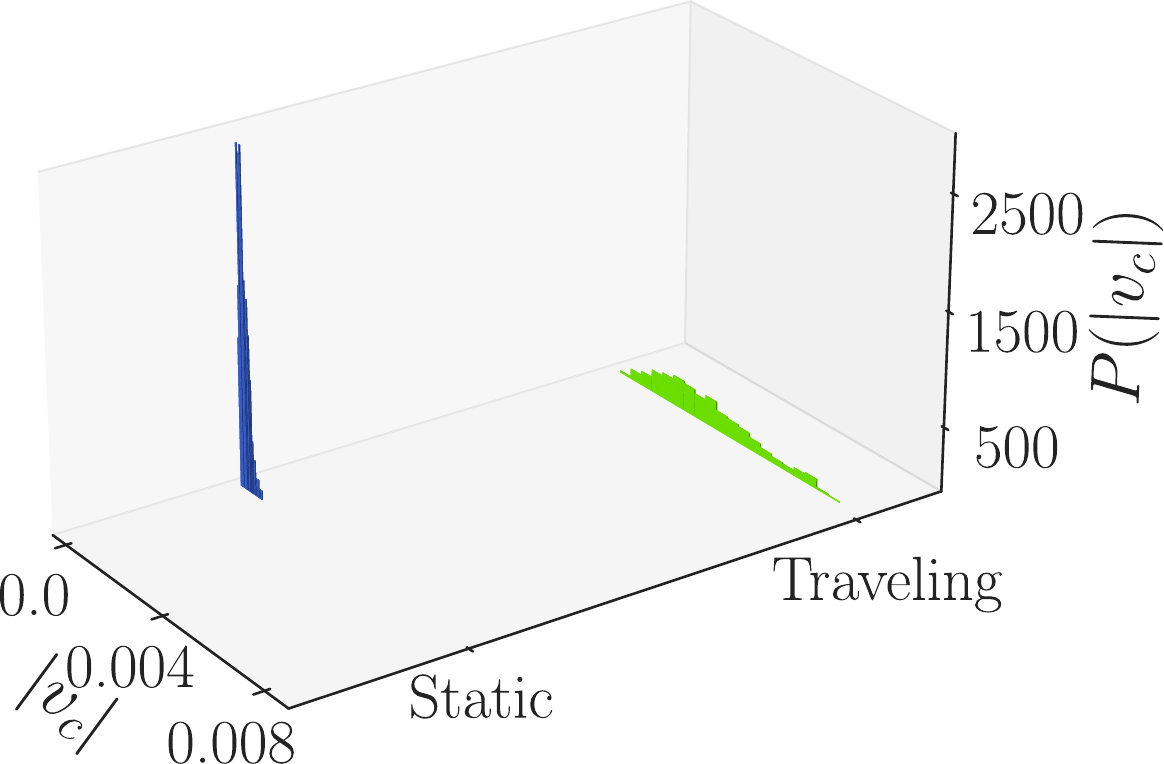}
\caption{Distributions of the magnitude of the linear velocity of the center of mass for the systems in a two-dimensional periodic box. 
Parameter values for the travelling and static states are as in Fig.~S4.
}
\label{fig:state_dist_pbc}
\end{figure}

\section{Phase separation order parameter}
\label{sec:phase}

\noindent To quantify the degree of spatial separation of both mixture components, the local segregation parameter defined in Ref.~\cite{brito2009} as
\begin{equation}
    \sigma = 1- \frac{\int d\mathbf{r} \;\rho_A(\mathbf{r})\rho_B(\mathbf{r}) }{\sqrt{ \int d\mathbf{r}\;\rho_A^2(\mathbf{r}) \int d\mathbf{r}\;\rho^2_B(\mathbf{r}) }},
\end{equation}\label{segregation}
was used. Here, $\rho_A(\mathbf{r})$ and $\rho_B(\mathbf{r})$ correspond to the local density of the A and B species, respectively. The values of $\sigma$ range from 1, for complete segregation, to zero for a maximally mixed system. The integral was implemented in the numerics by discretizing the simulation box into two dimensional square cells. Then, the contributions to the density from each individual cells were summed to calculate $\sigma$. As explained above we average $\sigma$ over time (in the steady state) and also over an independent ensemble to calculate $\langle \sigma \rangle$. 

\section{Switching Times}

\begin{figure}[ht]
\includegraphics[height=.65\linewidth]{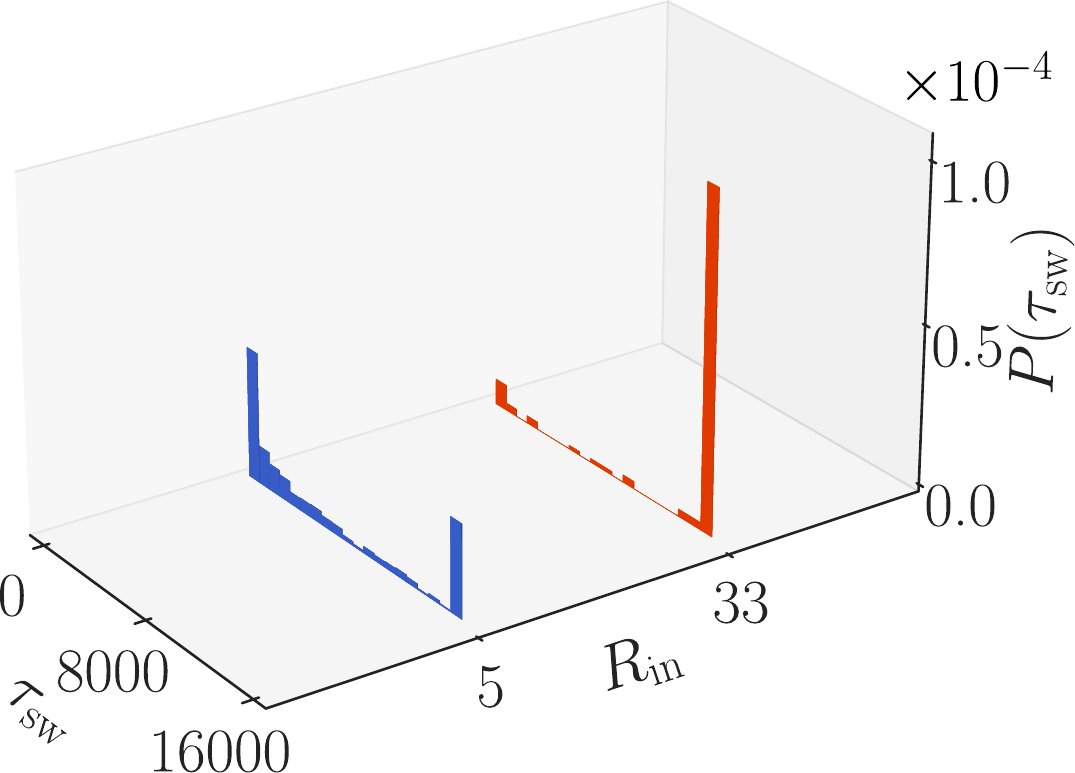}
\caption{Distribution of switching times of the solid type particles for two different values of the inner radius $R_{\rm in}$ shown in Fig.~3 of the main text.}
\label{fig:switch_times}
\end{figure}

\noindent We measure switching time to quantify the persistent motion of the travelling clusters. Fig.~S\ref{fig:switch_times} shows the distribution of switching time $P(\tau_{\rm SW})$ for particles within the solid cluster (species A for the chosen interaction parameters) for two different values of the inner radius ($R_{\rm in}=5, 33$). The switching time $\tau_{\rm SW}$  is defined as the time it takes for a single particle to go from rotating in a clockwise direction to an anticlockwise direction or vice-versa. This was estimated from the rotational velocity evaluated over a relatively long time interval $\Delta t=750$, in order to eliminate any diffusive fluctuations appearing at small time scales. In order to obtain the values at the steady state, data for an initial period (of duration $t_{\rm max}/5$, shorter than elsewhere to improve the statistics) were excluded from the calculation of $\tau_{\rm SW}$. Due to the finite run time of the simulation, we assign switching time equal to the possible maximum ($80\%$ of the total simulation time interval) if no change in rotation direction takes place in the steady state. Fig.~S\ref{fig:switch_times} shows that typical switching times for $R_{\rm in}=5$ are much smaller than the case for $R_{\rm in}=33$ in which the upper limit is much more affected by $t_{\rm max}$. 

\begin{figure}[ht]
\includegraphics[height =0.6\linewidth]{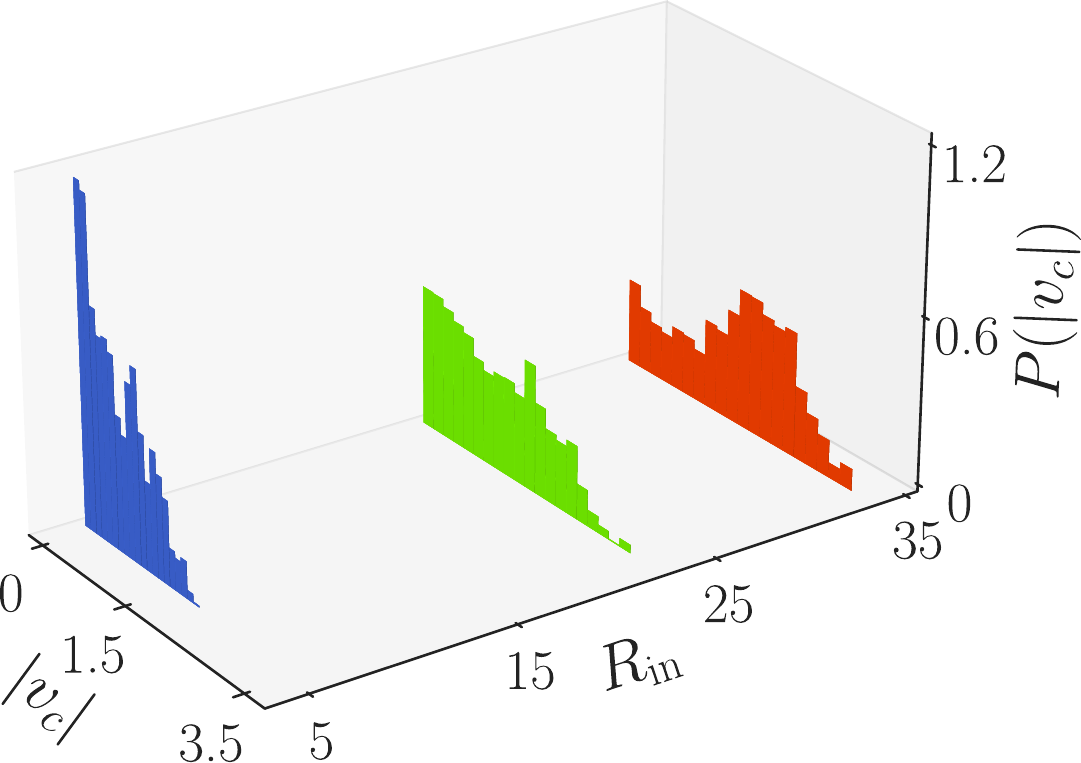}
\caption{Distribution of the magnitude of the linear velocity of the center of mass of the system, for different values of $R_{\rm in}$. For $R_{\rm in}=33$ a clear peak at finite $|v_c|$ is visible, which is a signature of the travelling state.}
\label{fig:v_dist}
\end{figure}

\section{Strong Non-reciprocity Limit (Predator-Prey)}

\begin{figure}
    \centering
    \includegraphics[height =0.9\linewidth]{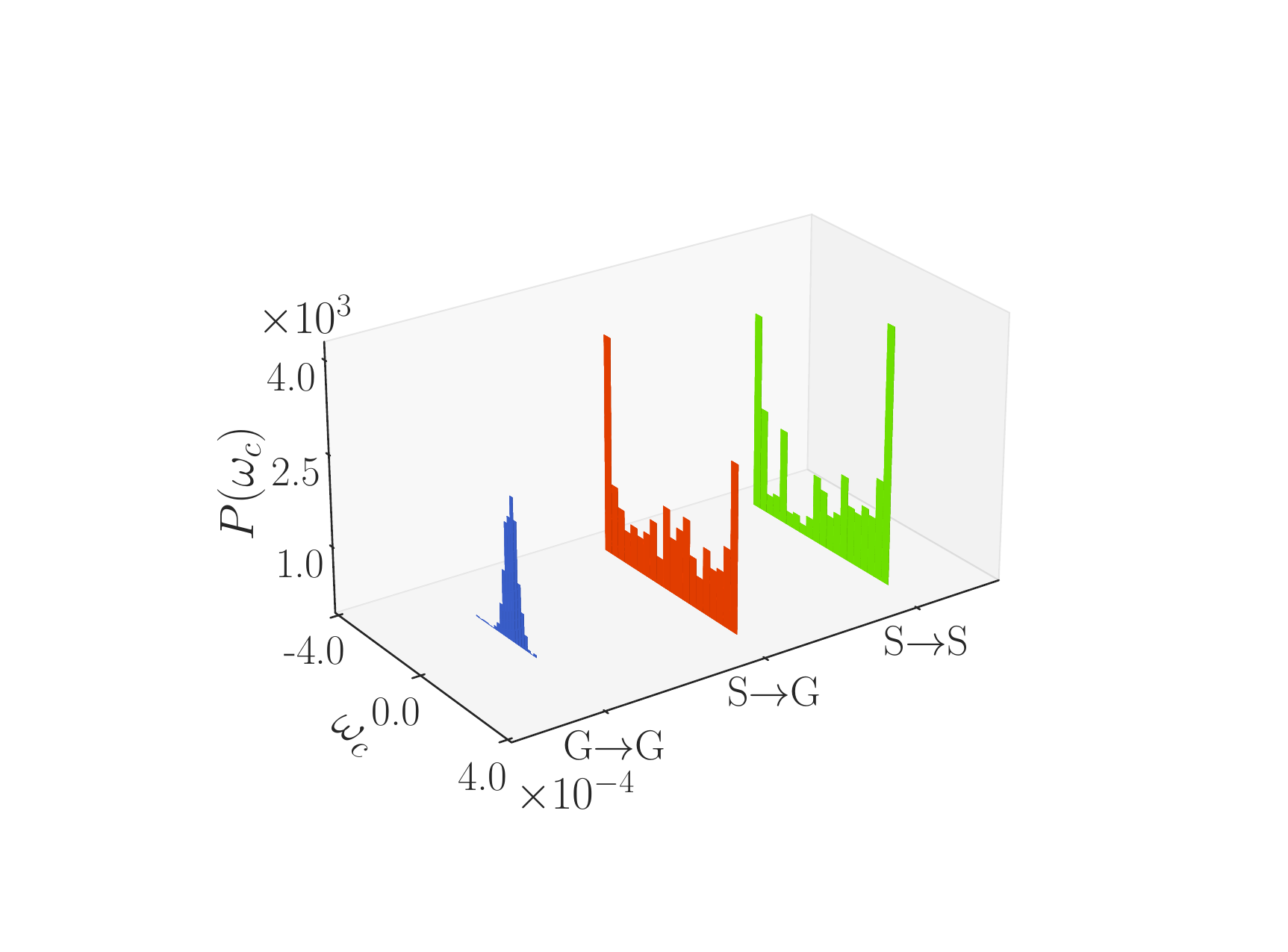}
    \caption{Probability distribution of centre of mass angular velocity $\omega_c$, for three combinations of aggregate states with parameters as in Fig.1 of the main text but $\delta=1.5$. The distribution for G $\to$ G has been scaled down by a factor of 4 for better visualization.}
    \label{fig:fig_S08}
\end{figure}

In this section we present how some of the results shown in the main text are also observed in systems with a higher level on non-reciprocity ({\it{e.g.}} $\delta \sim 1.5$: the predator-prey regime). Fig.S~\ref{fig:fig_S08} shows the distributions of the center of mass angular velocity $\omega_c$ for the mixture states of gas chasing gas $\rm G\rightarrow \rm G$, solid chasing gas $\rm S\rightarrow \rm G$ and solid chasing solid $\rm S\rightarrow \rm S$ in an annular confinement. Here the distributions for solid chasing solid and solid chasing gas has a clear bimodal shape, which is a feature of traveling states. On the other hand, the case with gas chasing gas is centered around zero, indicating that the state is static. This behaviour is consistent with the observation made in the main text, that traveling state are more commonly found when the chasing species is a solid, especially if the chased species is a gas (see Movies in the last section). 

\begin{figure}
    \centering
    \includegraphics[height =0.8\linewidth]{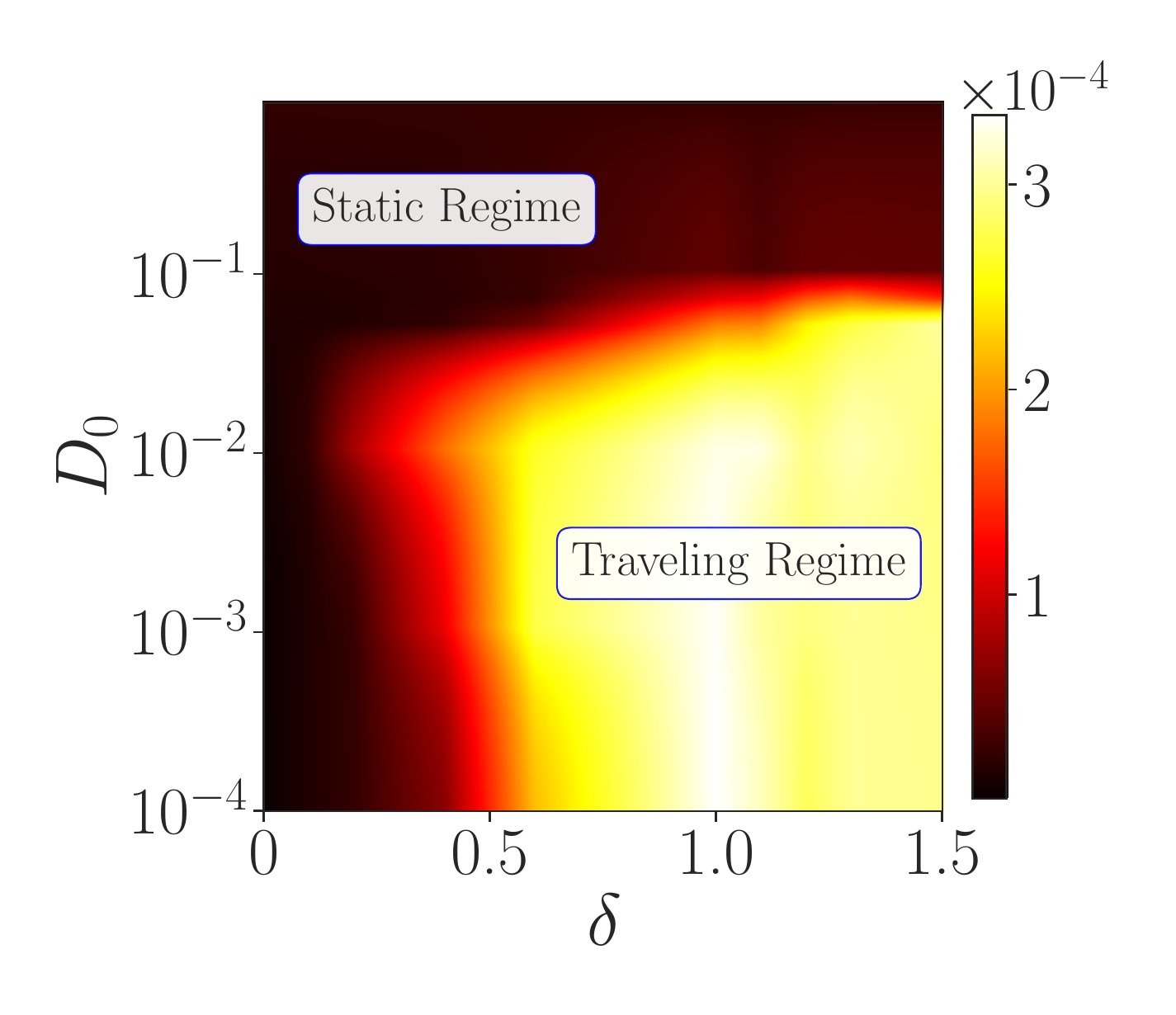}
    \caption{Extended phase diagram from Fig.2(e) of the main text, in the diffusivity and non-reciprocity plane, with data up to $\delta=1.5$, where the color represents the average absolute angular velocity $\langle |\omega|\rangle$.}
    \label{fig:fig_S09}
\end{figure}

Furthermore, to get a travelling state diffusivity needs to be small enough even at higher $\delta$'s, as can be seen from the extended $D_0-\delta$ phase diagram shown in Fig.S~\ref{fig:fig_S09}. Here it is evident that, when the diffusivity approaches a value of $0.1$, the average absolute angular velocity drops dramatically, a behaviour even true as $\delta$ goes above 1.0 (predator-prey regime).

\section{Finite size effects}

In this section we show how the results presented in the main text depend on the number of particles. In order to preserve the overall geometry of the annular confinement and maintain the same density of particles, we to increase the inner radius and $\Delta R$ as the number of particle increases. Following this scheme we preserve the "aspect ratio" of the system. 

\begin{figure}
\centering
\includegraphics[height =0.65\linewidth]{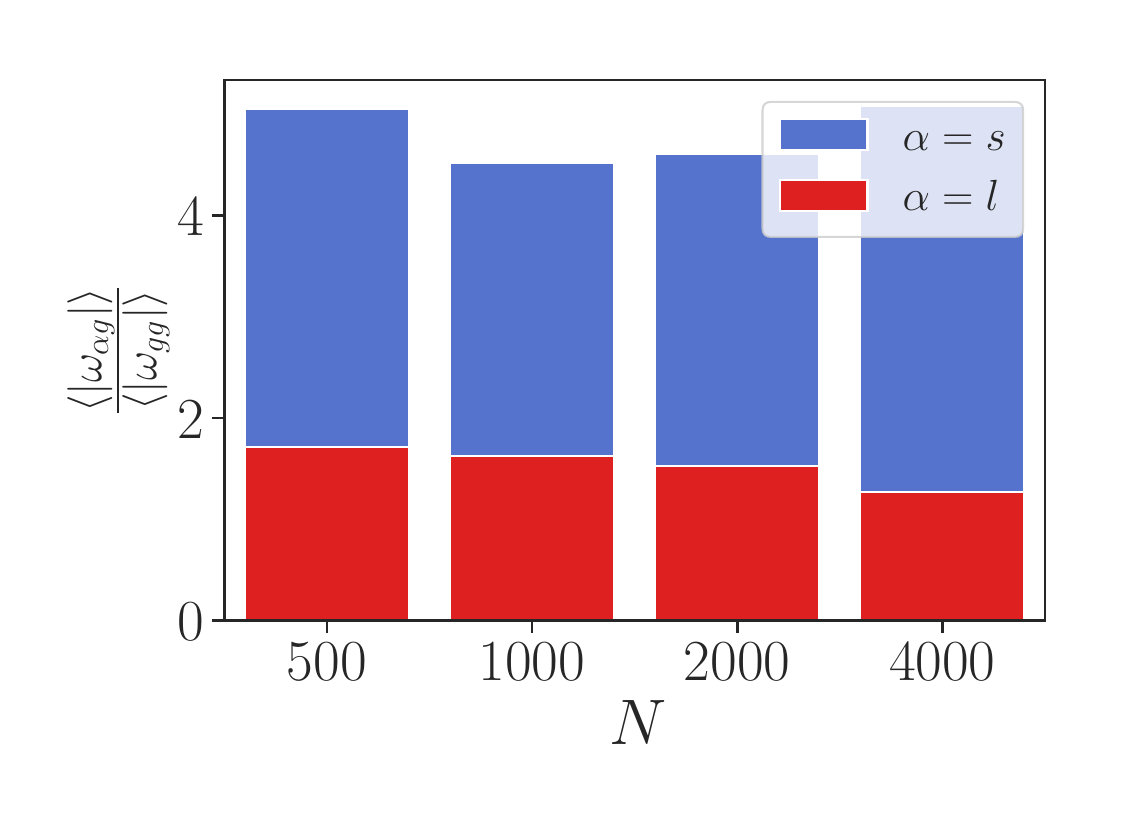}
\caption{Ratio between the angular velocity of a mixture with components having state (solid, liquid or gas) for different system sizes. Here $\langle |\omega_{\alpha g|} \rangle$ represents the angular speed of the mixture where $\alpha$ can represent either liquid (l) or solid (s) and a mixture of a gas chasing gas case ($\langle |\omega_{gg}| \rangle$). From the figure it is evident that a mixture of solid chasing gas ($\langle | \omega_{s g} |\rangle$) has a larger angular velocity than the liquid chasing gas ($\langle | \omega_{l g} |\rangle$) case consistently, even as system sizes grow. All systems have been kept at constant $\rho =0.1$, $D_0 = 0.01$ and $\delta = 0.9$} 
\label{fig:fig_S10}
\end{figure}

In Fig.S~\ref{fig:fig_S10}  we show that the solid chasing gas is still better candidate to observe the travelling state for larger system size than {\it{e.g.}} a mixture where the liquid component is chasing the gaseous one. Note we have normalised each case with corresponding mean velocity obtained for the gas chasing gas case, as we observe a systematic decrease in the chasing velocity. While increasing the system size we kept the aspect ratio of the geometry fixed. In this situation the travelling cluster grows with number of particles ($\sim N$) but number of particles in the boundary between two antagonistic species only grow as $\sqrt{(N)}$, leading to a overall reduction in the velocity of the travelling state with increasing $N$. 

\begin{figure}
\centering
\includegraphics[height =0.7\linewidth]{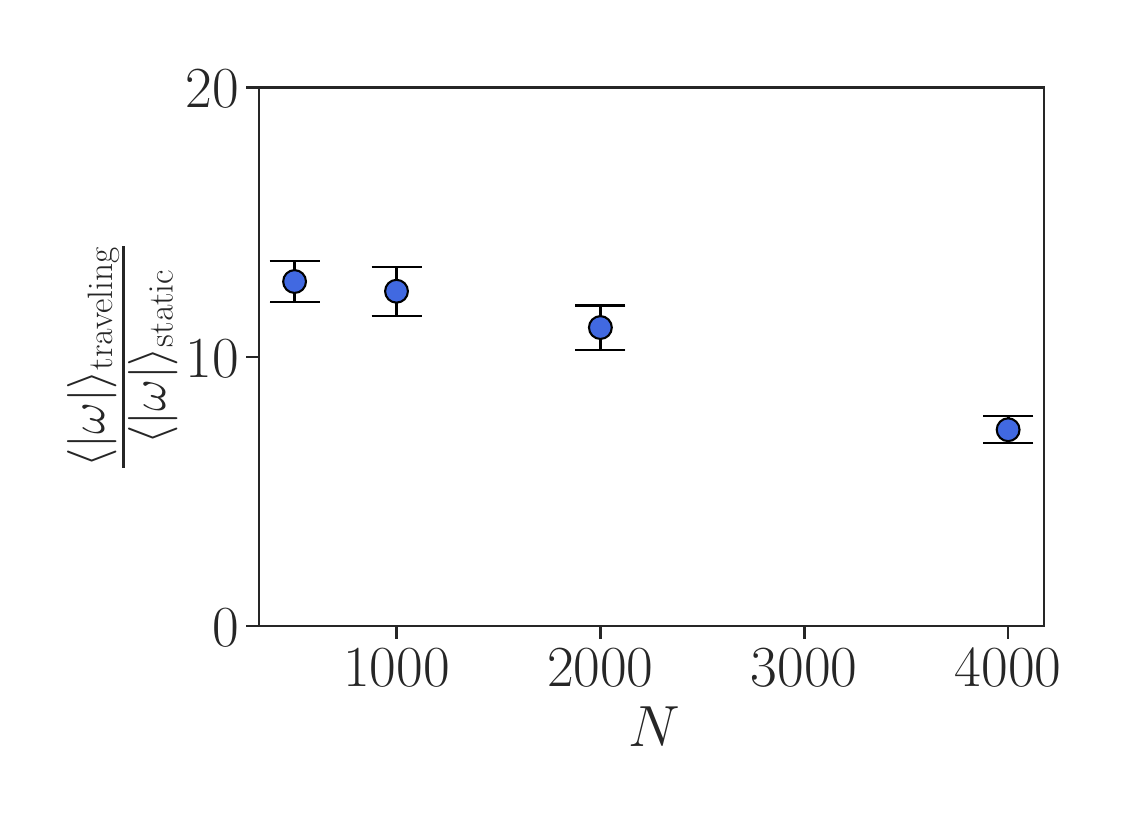}
\caption{Ratio of the mean absolute angular velocity of a traveling state ($\delta = 0.9 , D_0=10^{-3}$) and a static state ($\delta = 0.1 , D_0=0.1$) as a function of system size $N$ at constant $\rho=0.1$. Here it is evident, that states taken from the identified traveling regime have a significantly higher velocity than states taken from the static one even, at larger particle numbers.}
\label{fig:fig_S11}
\end{figure}

In Fig.S~\ref{fig:fig_S11} we show that states (traveling and static) chosen from the $\delta - D_0$ phase diagram given in the Fig.2 of the MS, (Fig.S~\ref{fig:fig_S09}), are still recognised as traveling and static states, even at larger system size $N$ grows. In this figure we have normalized the angular velocity of the traveling state by dividing it with the angular velocity of the static state. This is done to remove the trivial decrease (as mentioned in the previous paragraph) in angular velocity with increasing number of particles.

\vspace{0.5cm}

\vspace{0.5cm}

\section{Geometrical Effects}
 
  In order to prove that the bridging effect described in the main text is not due to any breaking of symmetry between inner or outer radius (or due to the difference in curvature), we ran simulations in a quasi-1D case. We first establish the equivalence between the quasi-1D geometry and annular geometry in Fig.S~\ref{fig:fig_S12} by comparing the linear velocities of the system (see supplementary movies that describe motion of travelling states in a quasi-1d geometry). Then we proceed with the quasi-1d geometry that described by periodic boundary conditions in the horizontal direction with size $L=80$ and fixed boundaries in the vertical direction with variable size $L_0$. Fig.S~\ref{fig:fig_S12} shows the correlation, between the speed of the travelling state and the number of `bridges' as we systematically vary the effective dimentionality of the system, that we also have observed in the annular geometry.

\begin{figure}
\includegraphics[width=0.99\linewidth]{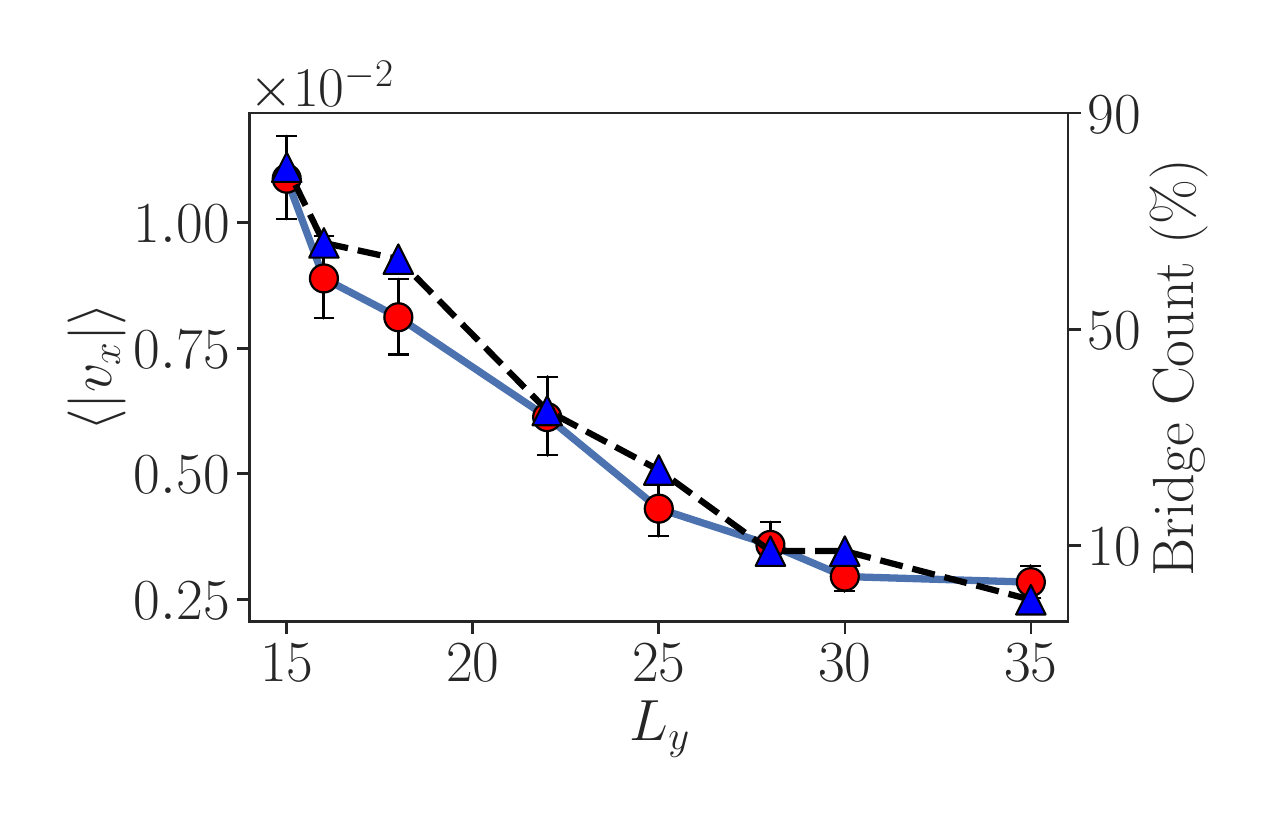}
\caption{Comparison of the average linear velocity of a traveling system confined to an annular geometry with $R_{\rm in} = 33$ and $R_{\rm out} = 42$ and it's direct equivalent in a quasi-1D geometry. The latter case has dimensions related to the annular case as follows: The y-axis length $L_{y} = \Delta R = 9.$, while the x dimension is $L_x = 2\pi (R_{\rm in} + R_{out})/2 = 235$. Both cases have the parameters $\rho=0.1$, $\delta = 0.9$ and $D_=0.01$.}
\label{fig:fig_S12}
\end{figure} 
\vspace{0.5cm}

\begin{figure}
\includegraphics[width=0.99\linewidth]{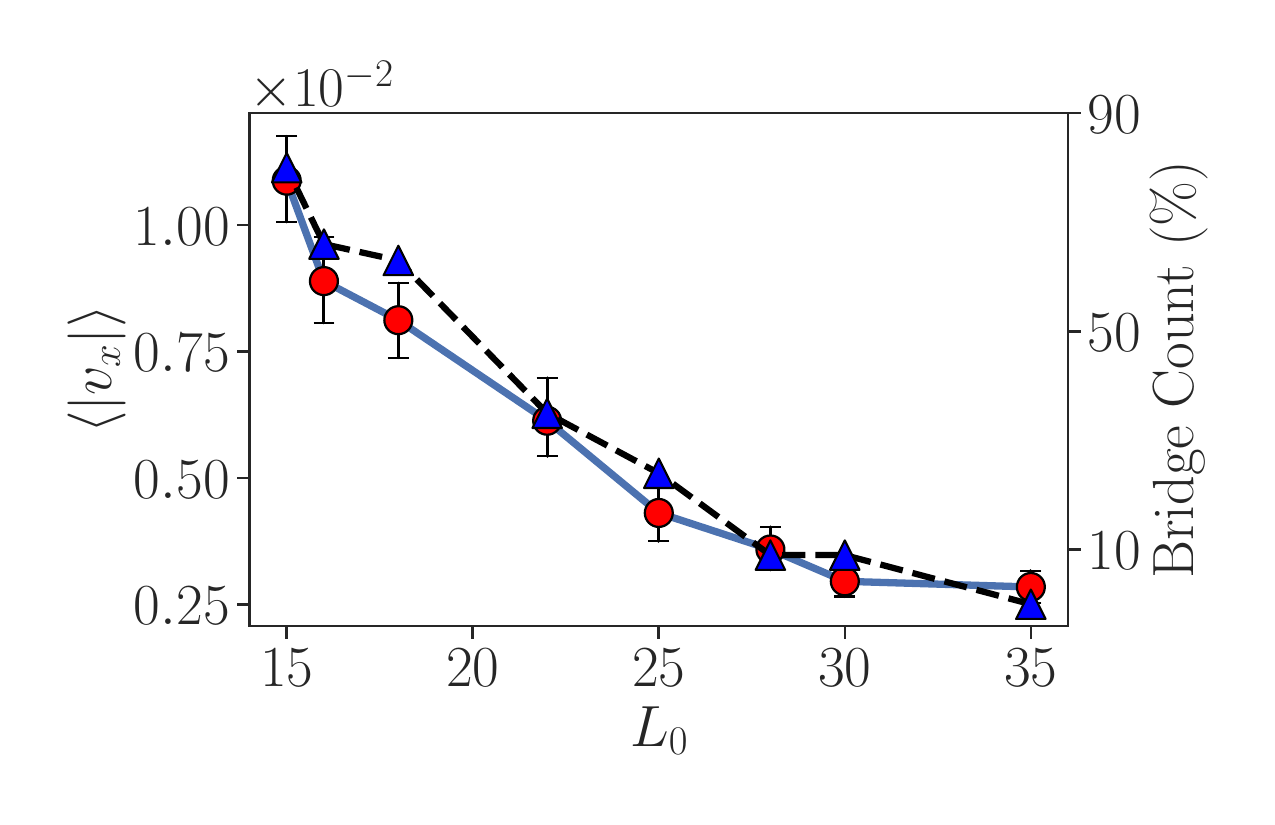}
\caption{Comparison between the average of the absolute value of the linear velocity along the horizontal direction $\langle |v_x| \rangle$ and number of bridges in a quasi-1D geometry plotted as a function of the channel width $L_0$, with channel length $L=80$. Here, periodic boundary conditions are used in the horizontal or x-direction and fixed boundaries in the vertical or y-direction. The triangles represent the number of bridges and the circles $\langle|v_x|\rangle$. The simulations where performed for $\delta = 0.9$, $D_0 = 0.01$ and $\rho=0.2$.}
\label{fig:fig_S13}
\end{figure} 
\vspace{0.5cm}

\section{Diffusivity and Density Phase Diagram}

\begin{figure}
\includegraphics[width=\linewidth]{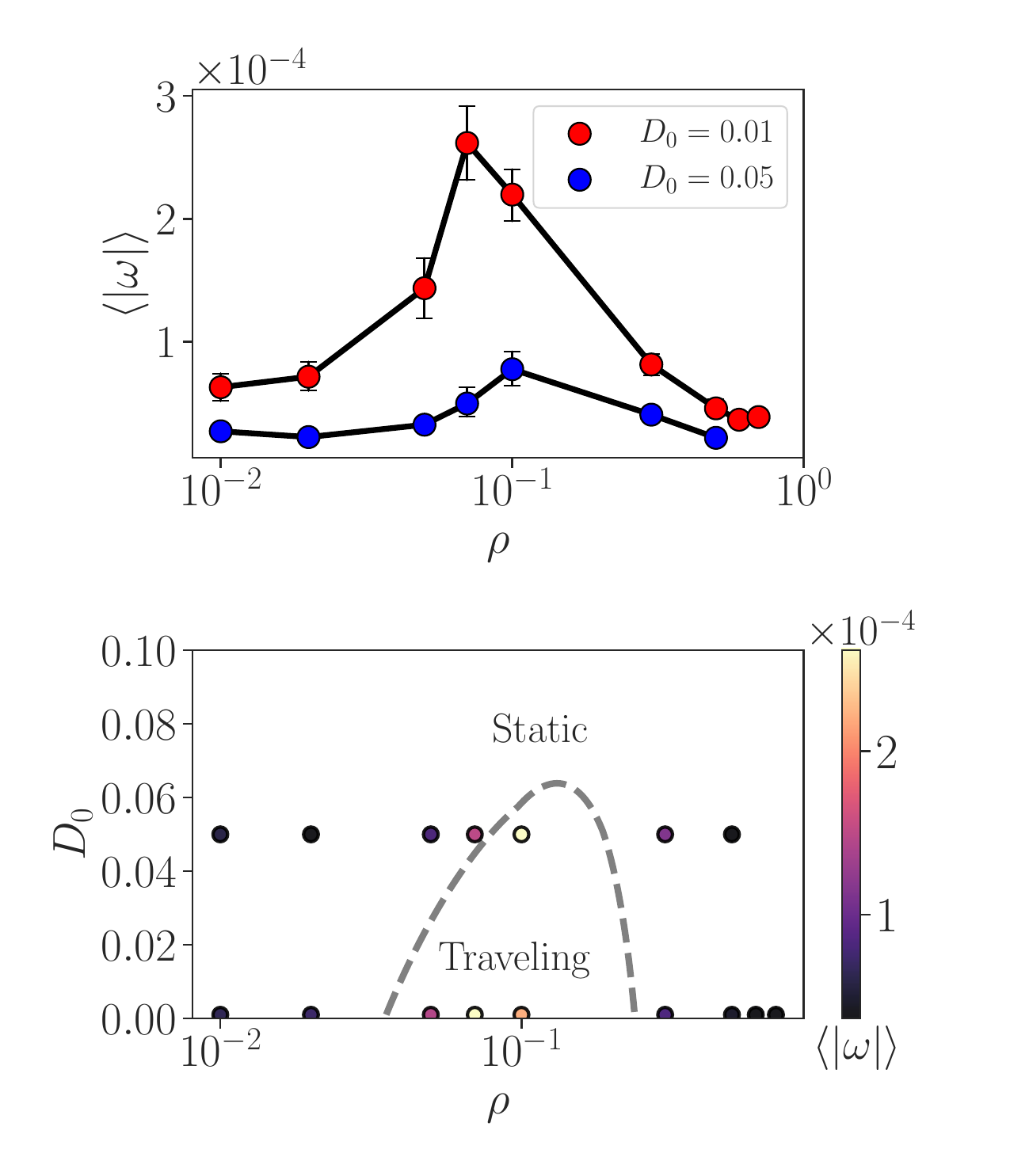}
\caption{Top plot: Average angular velocity $\langle |\omega| \rangle$ as a function of density $\rho$ for two fixed values of the diffusivity $D_0$. Bottom plot: Sketch of the phase diagram in the $D_0-\rho$ plane, based on the data from the left plot. The dots correspond to the data points shown on the left plot and the dashed line shows a plausible boundary between the travelling and the static phase.}
\label{fig:fig_S14}
\end{figure}

Here in this section, we tried to sketch our expectation of how the phase diagram would look like in the diffusivity-density plane, from our simulation data. Fig.S~\ref{fig:fig_S14} shows the dependence of the average angular velocity as a function of density for two different values of $D_0$ together with a sketch of the $D_0 -\rho$ phase diagram based on these data points. Our results suggest that traveling states are only possible in a relatively small region in the $D_0 - \rho$ plane. This region is situated towards the lower diffusivities and intermediate densities, where the system is not dominated by noise and has enough space to form a density gradient necessary to travel.

\vspace{0.5cm}

\section{Movies}


\begin{itemize}
    \item In the following link supplementary movie has been shared which describes that the particles from the gas phase can ``leak'' past the chasing cluster and this lead to the reversal of the direction of motion of the cluster. The movie has been generated for $R_{\rm in}=22$, $D_0=0.01$, $\rho=0.1$, $V_{\rm AA}=1.0$, $V_{\rm BB}=0.0$. Link: \url{https://github.com/Ssalazar12/Physics/blob/main/NonReciprical_Sup/Movie.mp4}
\item The following movies show how different combinations of the aggregate mixture states behave in the high non-reciprocity regime ($\delta = 1.5$):

\begin{enumerate}
    \item Solid chasing Gas \url{https://github.com/Ssalazar12/Physics/blob/main/NonReciprical_Sup/R1_movie1.mp4}.
    \item Solid chasing solid \url{https://github.com/Ssalazar12/Physics/blob/main/NonReciprical_Sup/R1_movie2.mp4}.
    \item Gas chasing gas \url{https://github.com/Ssalazar12/Physics/blob/main/NonReciprical_Sup/R1_movie3.mp4}.
\end{enumerate}

\item The following movies show the effect of increasing channel width $L_0$ in the quasi 1-D geometry:

\begin{enumerate}
    \item $L_0 = 15$ \url{https://github.com/Ssalazar12/Physics/blob/main/NonReciprical_Sup/R6_movie1.mp4}.
    
    \item $L_0 = 22$ \url{https://github.com/Ssalazar12/Physics/blob/main/NonReciprical_Sup/R6_movie2.mp4}.
    
     \item $L_0 = 35$ \url{https://github.com/Ssalazar12/Physics/blob/main/NonReciprical_Sup/R6_movie3.mp4}
\end{enumerate}
 
\item The following movie compares a system in annular confinement with $R_{\rm in}= 33$ and $R_{\rm out} = 42$, with its quasi-1D equivalent of dimensions $L_y = R_{\rm out} -  R_{\rm in}$ and $L_y = (R_{\rm out}+R_{\rm in})/2$. Both systems have $\rho = 0.1$, $D_0=0.01$ and $\delta = 0.9$. Link: \url{https://github.com/Ssalazar12/Physics/blob/main/NonReciprical_Sup/R7_movie.mp4}

\end{itemize}

\bibliography{si}
\bibliographystyle{ieeetr}